\documentclass[12pt]{article}

\usepackage{amsmath,amsfonts,amssymb}
\usepackage{amscd}
\usepackage{}
\usepackage{color}
\usepackage{comment}
\usepackage[utf8]{inputenc}

\newlength{\minitwocolumn}
\newcommand\bcQ{\mbox{\boldmath $Q$}}

\newcommand\tila{{\tilde{a}}}

\newcommand{\he}{\hat{\epsilon}}

\newcommand{\hdel}{\hat{\delta}}
\newcommand{\hFA}{\hat{F_A}}

\newcommand\bcx{\mbox{\boldmath $x$}}
\newcommand\bcy{\mbox{\boldmath $y$}}
\newcommand\bcA{\mbox{\boldmath $A$}}

\newcommand\bcC{\mbox{\boldmath $C$}}

\newcommand\bcXi{\mbox{\boldmath $\Xi$}}

\newcommand\bcd{\mbox{\boldmath $d$}}

\newcommand\bcZ{\mbox{\boldmath $Z$}}

\newcommand\bcF{\mbox{\boldmath $F$}}
\newcommand\bcH{\mbox{\boldmath $H$}}

\newcommand\bcX{\mbox{\boldmath $X$}}

\def\bbd{\mbox{\boldmath $d$}}

\begin{document}
\begin{titlepage}
\begin{flushright}
\null \hfill Preprint TU-1035\\[3em]
\end{flushright}

\begin{center}
{\Large \bf
Off-Shell Covariantization of Algebroid Gauge Theories
}
\vskip 1.2cm

Ursula Carow-Watamura${}^{a,}$\footnote{E-mail:\
ursula@tuhep.phys.tohoku.ac.jp}, Marc Andre Heller${}^{a,}$\footnote{E-mail:\
heller@tuhep.phys.tohoku.ac.jp}, Noriaki Ikeda${}^{b,}$\footnote{E-mail:\
nikeda@se.ritsumei.ac.jp}, Tomokazu Kaneko${}^{a,}$\footnote{E-mail:\
t\textunderscore kaneko@tuhep.phys.tohoku.ac.jp
} 
~and Satoshi Watamura${}^{a,}$\footnote{E-mail:\ watamura@tuhep.phys.tohoku.ac.jp}

\vskip 0.4cm
{

\it
${}^a$
Particle Theory and Cosmology Group, \\
Department of Physics, Graduate School of Science, \\
Tohoku University \\
Aoba-ku, Sendai 980-8578, Japan \\ 

\vskip 0.4cm
${}^b$
Department of Mathematical Sciences,
Ritsumeikan University \\
Kusatsu, Shiga 525-8577, Japan \\

}
\vskip 0.4cm



\begin{abstract}
We present a generalized method to construct field strengths and gauge symmetries, 
which yield a Yang-Mills type action with Lie n-algebroid gauge symmetry. The procedure 
makes use of off-shell covariantization in a supergeometric setting.
We apply this method to the system of a 1-form gauge field and scalar fields 
with Lie n-algebroid gauge symmetry. We work out some characteristic examples.

\end{abstract}
\end{center}
\end{titlepage}

\newpage

\setcounter{tocdepth}{2}

\section{Introduction}

Recently, many approaches for a generalization of gauge theories
are being discussed. Among them, there are the so-called higher gauge theories \cite{Baez:2002jn}, where in addition to the 
gauge potential higher rank forms are introduced.
Such theories are expected to appear, for example, in the construction of  
the effective theory of multiple M5-branes where a 2-form gauge potential appears. 

Another approach is the promotion of the gauge algebra to an 
algebroid structure. This can be thought of as a generalization of the gauged non-linear sigma model,  
where the structure constants of the Lie algebra become scalar field dependent
\cite{Ikeda:2001fq, Strobl:2004im, Kotov:2015nuz}.

A systematic way to construct higher gauge theories
is to use an $L_\infty$-structure \cite{Lada:1992wc}. Any truncated $L_\infty$-algebra 
defines a gauge theory of higher form gauge fields and the
corresponding gauge symmetries are generalized to Lie n-algebras \cite{Baez:2003fs}.
As we shall see, both generalizations, i.e., to higher gauge theory and to algebroid Yang-Mills can be understood in a unified way using supergeometry.  
Actually, there is a common phenomenon in both approaches, i.e. the higher gauge theory using 
an $L_\infty$-structure and the approach via algebroid structure, when it comes to the formulation of the corresponding field theories.
This phenomenon is the so-called fake curvature condition \cite{Baez:2010ya}. 

In a general higher gauge theory also lower form gauge fields exist. 
However, the field strength of the higher form gauge potential is only covariant under the condition that the field strengths associated to the lower form gauge fields vanish. This is called the fake curvature condition and results in a non-interactive theory. 
Therefore, it is desirable to deform the higher algebra structure to circumvent this obstruction.
Such a deformation process, known as \emph{off-shell covariantization}, has been analyzed in the higher gauge theory context in our previous paper \cite{Carow-Watamura:2016nwj}. There, we solved the fake curvature 
condition by reducing the symmetries to Lie n-subalgebras, while imposing proper conditions on the auxiliary gauge fields.

In this paper, we want to address the problem of off-shell covariantization in the context of algebroid gauge theories. 
We apply our method to systems consisting of a 1-form gauge field and 
a scalar field. We formulate the corresponding higher algebroid gauge symmetries and associated gauge invariant actions.
To obtain off-shell Yang-Mills type actions,
we consider deformations of gauge transformations and field strengths.
Auxiliary gauge fields are projected out and
field strengths are deformed by terms proportional to the lower curvatures.

In order to obtain proper gauge symmetries of gauge fields and field strengths, we use 
the supermanifold method on a so-called QP-manifold \cite{Schwarz:1992nx,Schwarz:1992gs}, which is a useful tool to generate a BRST-BV formalism
of topological field theories \cite{Alexandrov:1995kv}. 
Instead of starting from fields and an action, 
we start with a graded symplectic manifold 
and its Hamiltonian function corresponding to a BRST charge of the gauge algebra.
Gauge fields, field strengths and their gauge transformations 
are induced from the QP-manifold structure.
This idea is similar to the free differential algebra method 
\cite{Sullivan, D'Auria:1982nx}.
In our formalism, consistency is guaranteed by the underlying 
QP-manifold structure
\cite{Alexandrov:1995kv,Fiorenza:2010mh,Gruetzmann:2014ica,Lavau:2014iva}.

The advantage of the supermanifold method is that the gauge transformations
and field strengths can be derived in a systematic manner.
The starting point of our analysis is a general theory unifying gauge theories with algebroid
symmetry and those with Lie n-algebra symmetry.
Examples are the Kotov-Strobl model \cite{Kotov:2015nuz} 
and the Ho-Matsuo model \cite{Ho:2012nt}.
See also \cite{Strobl:2016aph} for a gauge theory with 
a Lie 2-algebra symmetry.

The organization of this paper is as follows. In section 2, 
we briefly review QP-manifolds and explain the off-shell covariantization procedure
used in this paper.
In section 3, 
we discuss the construction of $(n+1)$-dimensional higher algebroid gauge theories 
based on general QP-manifold structures.
In section 4, we construct and analyze 4-dimensional algebroid gauge theories.
We derive the relations between the structure functions necessary for
off-shell covariantization. Furthermore, we discuss examples
including the St\"uckelberg formalism, non-abelian off-shell covariantization
and an example from the Kotov-Strobl models.
In section 5, we examine the closure of the gauge symmetry algebra.
Section 6 is devoted to discussion.

\section{QP-manifolds and off-shell covariantization}

In this section, we briefly review how to construct 
gauge transformations and field strengths using QP-manifolds.
Then, we shortly explain the off-shell covariantization
procedure of field strengths. Please refer to \cite{Ikeda:2012pv, Bessho:2015tkk} for conventional details.

A QP-manifold ($\mathcal{M}$, $\omega$, $\bcQ$) of degree $n$ consists of a nonnegatively graded manifold $\mathcal{M}$, a symplectic structure $\omega$ of degree $n$ and a homological vector field $\bcQ$ of degree $1$ on $\mathcal{M}$ such that $L_{\bcQ} \omega = 0$. The requirement of $\bcQ$ to be homological is equivalent to saying $\bcQ$ is nilpotent, $\bcQ^2 = 0$. The graded symplectic structure induces a Poisson bracket $\{-,-\}$ of degree $(-n)$. 	

For any QP-manifold one can find a function $\Theta\in\mathcal{C}^\infty(\mathcal{M})$ of degree $n+1$, such that
\begin{equation}
	\bcQ = \{\Theta,-\}.
\end{equation}
The nilpotency of $Q$ then translates to the \emph{classical master equation}
\begin{equation}
	\{\Theta, \Theta\} = 0.
\end{equation}
A QP-manifold can also be called a symplectic NQ-manifold. The operator $\bcQ$ generates BRST transformations of the associated gauge theory.

Though the method can be used to construct general $p$-form gauge theories, in this paper 
we focus on a theory containing scalar fields $X^i(\sigma)$ 
and a 1-form gauge field $A^a = d \sigma^{\mu} A_{\mu}^a(\sigma)$.

Our set-up is as follows. We consider a QP-manifold of degree $n$, where the graded manifold is given by $\mathcal{M}_n = T^*[n]E[1]$\footnote{$[1]$ and $[n]$ denote shifting of degree by 1 and $n$, respectively.} , $n\in\mathbb{N}$, and $E\rightarrow M$ is a vector bundle. $M$ is a smooth manifold. We take the following local coordinates: $x^i$ of degree $0$ on $M$  and $q^a$ of degree $1$ on the fiber of the vector bundle. When we construct the associated field theory, the degree corresponds to the ghost degree. With respect to the graded cotangent bundle $T^*[n]$, we take coordinates ($\xi_i$, $p_a$) of degree ($n$, $n-1$) conjugate to ($x^i$, $q^a$). To summarize, the local coordinates on $\mathcal{M}_n$ are ($x^i$, $q^a$, $\xi_i$, $p_a$) of degree ($0$, $1$, $n$, $n-1$).

The symplectic form on $\mathcal{M}_n$ is defined by
\begin{equation}
	\omega = \delta x^i \wedge \delta \xi_i + (-1)^n \delta q^a \wedge \delta p_a.
\end{equation}
This induces the following graded Poisson bracket,
\begin{equation}
\{f,g\}
= \frac{f\overleftarrow{\partial}}{\partial x^i}\frac{\overrightarrow{\partial} g}{\partial \xi_i}-\frac{f\overleftarrow{\partial}}{\partial \xi_i}\frac{\overrightarrow{\partial} g}{\partial x^i}+
\frac{f\overleftarrow{\partial}}{\partial q^{a}}\frac{\overrightarrow{\partial} g}{\partial p_{a}}+(-1)^{n}\frac{f\overleftarrow{\partial}}{\partial p_{a}}\frac{\overrightarrow{\partial} g}{\partial q^{a}},
\end{equation}
where $f,g\in\mathcal{C}^\infty(\mathcal{M}_n)$.\footnote{The relation between right and left derivative is given by $\frac{f\overleftarrow{\partial}}{\partial X}=(-1)^{|X|(|f|-|X|)}\frac{\overrightarrow{\partial} f}{\partial X}$.}

By making use of the BV-ASKZ formalism \cite{Alexandrov:1995kv, Ikeda:2012pv}, a topological field theory in $n + 1$ dimensions can be constructed from a QP-manifold of degree $n$. Let $\Sigma$ be the worldvolume. The starting point of the construction is the promotion of the worldvolume to a graded space space $T[1]\Sigma$. We denote the local coordinates of $\Sigma$ by $\sigma^\mu$, which are of degree $0$, and those of the fiber by $\theta^\mu$, which are of degree $1$.

Let $\mathcal{M}_n$ be our QP-manifold. Then we can define the map $a: T[1]\Sigma \rightarrow {\cal M}_n$, such that \cite{Gruetzmann:2014ica}
\begin{equation}
\bcZ(\sigma, \theta) \equiv a^{*}(z) = \sum_{j=0}^{n+1}\bcZ^{(j)}(\sigma, \theta) = \sum_{j=0}^{n+1} \frac{1}{j!} \theta^{\mu_1} \cdots \theta^{\mu_j} Z^{(j)}_{\mu_1 \cdots \mu_j}(\sigma)
\end{equation}
is a superfield. Here, $z$ is a coordinate of degree $k$ on $\mathcal{M}_n$. The map $a$ is degree-preserving so that $|\bcZ| = k$. Since the resulting object is a superfield in the BV sense, it contains associated gauge fields, ghosts and antifields as component fields. The physical component is the ghost number $0$ component. In general, the ghost number of a field $\Psi$ is defined by degree minus form degree, $\mathrm{gh}(\Psi) = |\Psi| - \mathrm{deg}(\Psi)$, where form degree $(0,1)$ is assigned to $(\sigma^{\mu}, \theta^{\mu})$.

By degree counting, $Z^{(j)}_{\mu_1 \cdots \mu_j}$ has ghost number $(k-j)$, $Z^{(k)}_{\mu_1 \cdots \mu_k}$ has ghost number $0$ and $Z^{(k-1)}_{\mu_1 \cdots \mu_{k-1}}$ has ghost number $1$. The ghost number $0$ component is a physical $k$-form gauge field and the ghost number $1$ component its FP ghost, i.e., the gauge parameter of the associated gauge transformation.

The \textit{super field strength} and its physical component corresponding to a 
coordinate $z$ on $\mathcal{M}$ are defined by 
\begin{align}
\bcF_Z &= \bcd \circ a^{*}(z) - a^{*} \circ \bcQ(z),
\label{superfieldstrength} \\
{\cal F}_z &= (\bcd \circ a^{*}(z) - a^{*} \circ \bcQ(z)) |_{|z|+1}.
\label{generalfieldstrength}
\end{align}
Here, $\bcd = \theta^{\mu} \partial_{\mu}$ denotes the superderivative and $|_{|z|+1}$ denotes projection to the degree $|z|+1$ part, while setting all antifield components to zero. We get the super Bianchi identity for free,
\begin{equation}
	(\bcd \circ a^{*}- a^{*} \circ \bcQ)^2=0 \; \Rightarrow \; dF_z=-F\circ\bcQ(z).
\label{Bianchi}
\end{equation}
The associated gauge transformation is encoded in the super field strength as the degree $|z|$ part,\footnote{This formula gives a BRST transformation.} 
\begin{align}
\delta Z = (\bcd \circ a^{*}(z) - a^{*} \circ \bcQ(z)) \big|_{|z|}.
\label{superfeildstrength}
\end{align}
Again, while projecting, we set all antifields to zero.

To extract the physical field directly, we define the map $\tila: T[1]\Sigma \rightarrow {\cal M}_n$ by
\begin{align}
\tila^{*}(z)& = \frac{1}{k!} d\sigma^{\mu_1} \wedge \cdots \wedge d\sigma^{\mu_k} 
Z^{(k)}_{\mu_1 \cdots \mu_k}(\sigma),
\label{expansionofsuperfield}
\end{align}
where $z$ is a coordinate of degree $k$ on $\mathcal{M}_n$. Note that we have identified $\theta^{\mu}$ with $d\sigma^{\mu}$. Using this map, we can rewrite the physical field strength by
\begin{equation}
F_z = F(z)  = d \circ \tila^{*}(z) - \tila^{*} \circ \bcQ(z).
\label{generalfieldstrength}
\end{equation}

For the QP-manifold $\mathcal{M}_n$ under consideration, we get a scalar field associated to the degree $0$ coordinate $x^i$ and a $1$-form gauge field associated to the degree $1$ coordinate $q^a$, and associated field strengths,
\begin{align}
\tila^{*}(x^i)&\equiv X^i(\sigma),\\
\tila^{*}(q^{a})&\equiv A^{a}(\sigma)=A^{a}_{\mu}d\sigma^{\mu}, \\
F_X^i &=d \tila^{*}(x^i)-\tila^{*}(\bcQ x^i),
\label{fieldstrengthFx} \\
F_A^{a} &= d \tila^{*}(q^{a})-\tila^{*}(\bcQ q^{a}).
\label{fieldstrengthFA}
\end{align}
In addition to that, we find $(n-1)$- and $n$-form auxiliary gauge fields $C_a$ and $\Xi_i$ associated with the conjugate coordinates on our graded symplectic manifold,
\begin{align}
\tila^{*}(x^i)& =X^i,\ \tila^{*}(q^a)=A^a,\\
\tila^{*}(\xi_i)& = \Xi_i, \ \tila^{*}(p_a)=C_a.
\end{align}
In this very scenario, we have gauge transformations with three independent gauge parameters corresponding to the fields $A^a$, $\Xi_i$ and $C_a$:
\begin{equation}
a^{*}(q^a)|_{\sc{deg}=0}=\epsilon^a, \quad a^{*}(\xi_i)|_{\sc{deg}=n-1} = \mu^{\prime}_i, \quad \tila^{*}(p_a)|_{\sc{deg}=n-2}=\epsilon^{\prime}_a.
\end{equation}
However, in general, the field strength $F_z$ of a gauge field 
$\tilde{Z} = \tila^*(z)$ is transformed adjointly, $\delta F_z \sim F_z$, 
only on-shell since the above procedure is derived from the theory of 
AKSZ sigma models \cite{Alexandrov:1995kv}.
An action of the AKSZ sigma models is a topological field theory of BF type
and the equation of motion is $\bcF_z=0$. If $F_z$ transforms adjointly without use of the equations of motion, we call
$F_z$  \textit{off-shell covariant}.
If $F_z$ is off-shell covariant, the construction of a gauge invariant 
Yang-Mills type action $S \sim F_z \wedge * F_z$ is possible.

The procedure to obtain off-shell covariant field strengths
is as follows.
First, we drop the auxiliary gauge fields $(\Xi_i, C_a)$
and extra gauge degrees of freedom $\mu^{\prime}_i$ and 
$\epsilon^{\prime}_a$
by the projection $\bcXi_i = \bcC_a =0$.
Then, we deform the field strengths $F_z$ and gauge symmetries $\delta$
by adding deformation terms proportional to lower field strengths.
Note that the algebra of the structure constants (or functions) 
is not deformed.
Choosing proper coefficients for the deformation leads to  
off-shell covariantized field strengths
without changing the original gauge symmetry algebra.
\section{Hamiltonian functions}
A Hamiltonian function $\Theta$ on a general QP-manifold ${\cal M}_n$ of degree $n$ is
of degree $n+1$.
In this section, we examine the most general Hamiltonian function 
on ${\cal M}_n$
by expanding it in conjugate coordinates $(\xi_i, p_a)$,
\begin{align}
\Theta &= \sum_{k}
\Theta^{(k)},
\label{expansionsofTheta}
\end{align}
where $\Theta^{(k)}$ is a $k$-th order function in $(\xi_i, p_a)$.

The following cases occur.
\paragraph{A)}
$n \geq 4$:
Since the degrees of $(\xi_i, p_a)$ are $(n, n-1)$,
the degree of $\Theta^{(k)}$ for $k \geq 2$ is 
larger than $2n-2$.
Therefore, if $n \geq 4$, then $\Theta^{(k)} = 0$ for $k \geq 2$ by 
degree counting, i.e. the general form of the Hamiltonian function is
\begin{align}
\Theta = \Theta^{(0)} + \Theta^{(1)}.
\end{align}
\paragraph{B)}
$n = 3$:
In this case, $\Theta^{(k)} = 0$ for $k \geq 3$ by degree counting.
Therefore, the expansion stops at second order,
\begin{align}
\Theta = \Theta^{(0)} + \Theta^{(1)} + \Theta^{(2)}.
\end{align}
this QP-manifold defines a Lie $3$-algebroid 
structure on $E$, see \cite{Ikeda:2010vz}.
Only for $n \leq 3$ the Hamiltonian $\Theta$ provides 
freedom for deformations. 
We discuss case $n=3$ in detail in section \ref{4d1form}.
\paragraph{C)}
$n=1, 2$:
The Hamiltonian $\Theta$ contains more deformation terms.
In the $n=2$ case,
since $(x^i, q^{a}, \xi_i, p_{a})$ is of degree $(0, 1, 2, 1)$,
the graded manifold is
${\cal M}_2 =  T^*[2]E[1]$, and 
\begin{align}
\Theta = \Theta^{(0)} + \Theta^{(1)} + \Theta^{(2)} + \Theta^{(3)}.
\end{align}
Then, this defines a Courant algebroid on $E$ \cite{Liu-Weinstein-Xu, Roytenberg:1999}.

For $n=1$, $\Theta$ defines a Poisson structure on $E$.

\subsection{Gauge fields and field strengths induced from Hamiltonian functions}
First, 
the Hamiltonian function $\Theta^{(1)}$ reproduces a
Lie algebroid for general $n$.
It contains the following terms,
\begin{align}
\Theta^{(1)}
&= \rho^i{}_a(x) \xi_i q^a
+ \frac{1}{2} f^c{}_{ab}(x) q^a q^b p_c,
\end{align}
where $\rho^i{}_a(x), f^c{}_{ab}(x)$
are structure functions depending on $x$.

Lie algebroid operations are given by
the following derived brackets,
\begin{align}
[e_1, e_2]&= - \{\{e_1,\Theta^{(1)} \}, e_2\}, 
\label{operation1}
\\
\rho(e) f &=\{\{e,\Theta^{(1)} \}, f\}, 
\label{operation2}
\end{align}
where $e, e_1, e_2 \in \Gamma(E)$ are sections of a Lie algebroid 
which is locally expressed by $e = e^a(x) p_a$
and $f \in C^{\infty}(M)$.

For details and notation, see appendix A.

The classical master equation, $\{\Theta^{(1)},\Theta^{(1)}\}=0$,
implies the following conditions on the structure constants,
\begin{eqnarray}
&&
\rho{}^{j}{}_a \frac{\partial \rho{}^{i}{}_b}{\partial x^j} 
- \rho{}^{j}{}_b \frac{\partial \rho{}^{i}{}_a}{\partial x^j} 
+ \rho{}^{i}{}_c f{}^c{}_{ab} = 0,
\\
&&
\rho{}^{j}{}_{[a} \frac{\partial f{}^d{}_{bc]}}{\partial x^j} 
+ f{}^d{}_{e[a} f{}^e{}_{bc]} = 0.
\label{Liealgebroid}
\end{eqnarray}

The pullback $a^*$ maps the four coordinates to superfields as follows,
\begin{align}
\bcX^{i} &\equiv a^{*}(x^i),\ \bcA^{a} \equiv a^{*}(q^{a}), \\
\bcXi_i & \equiv a^{*}(\xi_i), \ \bcC_a \equiv a^{*}(p_a).
\end{align}
The super field strengths are given by
$\bcF_Z = \bbd a^*(z) - a^* \bcQ (z)$.
\begin{align}
\bcF_X^{i}&= \bcd \bcx^i - \rho^i{}_a(\bcX) \bcA^a,
\label{3superfieldstrength1}
\\
\bcF_A^{a}&= \bcd \bcA^{a} + \frac{1}{2}f^{a}_{bc}(\bcX) \bcA^{b} \bcA^{c},
\label{3superfieldstrength2}
\\
\bcF^{(C)}_{a}&= \bcd \bcC_{a} + f^{b}_{ac}(\bcX) \bcC_b \bcA^c
- \rho^i{}_a(\bcX) \bcXi_{i},
\label{3superfieldstrength3}
\\
\bcF^{(\Xi)}_{i}&=\bcd \bcXi_i 
- \frac{1}{2} \frac{\partial f^a{}_{bc}}{\partial \bcx^i}(\bcX)
\bcC_a \bcA^{b} \bcA^{c} 
- \frac{\partial \rho^j{}_{a}}{\partial \bcx^i}(\bcX)\bcXi_j\bcA^{a},
\label{3superfieldstrength4}
\end{align}
where $\bcF^{(C)}$ and $\bcF^{(\Xi)}$ are the super field strengths of 
$\bcC$ and $\bcXi$, respectively.
When we substitute the component expansions to 
\eqref{3superfieldstrength1}--\eqref{3superfieldstrength4}, then the
corresponding degree $|z|+1$ parts are the field strengths:
\begin{align}
F_X^{i}&= d x^i - \rho^i{}_a(X) A^a,
\label{fieldstrength1}
\\
F_A^{a}&= d A^{a} + \frac{1}{2}f^{a}_{bc}(X) A^{b} A^{c},
\label{fieldstrength2}
\\
F^{(C)}_{a}&= d C_{a} + f^{b}_{ac}(X) C_b A^c
- \rho^i{}_a \Xi_{i},
\label{fieldstrength3}
\\
F^{(\Xi)}_{i}&= d \Xi_i 
- \frac{1}{2} \frac{\partial f^a{}_{bc}}{\partial X^i}(X)
C_a A^{b} A^{c} 
- \frac{\partial \rho^j{}_{a}}{\partial X^i}(X)
\Xi_j A^{a}.
\label{fieldstrength4}
\end{align} 
The degree $|z|$ parts of the component expansions of the super field strengths yield the
gauge transformations,
\begin{align}
\delta X^i &= - \rho^i{}_a(X) \epsilon^{a},
\\
\delta A^{a}&= d \epsilon^{a} + f^{a}_{bc}(X) A^{b} \epsilon^{c},
\\
\delta C_{a}&= d \epsilon^{\prime}_{a} 
+ f^{b}_{ac}(X) (\epsilon^{\prime}_{b} \wedge A^{c} 
+ C_{b} \wedge \epsilon^{c})
- \rho^i{}_a(X) \mu^{\prime}_i,
\\
\delta \Xi_{i}&= d \mu^{\prime}_{i} 
- \frac{1}{2} \frac{\partial f^a{}_{bc}}{\partial X^i}(X)
(\epsilon^{\prime}_a A^{b} A^{c} + 2 C_a A^{b} \epsilon^{c})
- \frac{\partial \rho^j{}_{a}}{\partial X^i}(X)
(\mu^{\prime}_j A^{a} + \Xi_j \epsilon^{a}).
\end{align}
The gauge transformations of the gauge field strengths are
\begin{align}
\delta F_X^{i} &= \frac{\partial \rho^{i}{}_a}{\partial X^j} F_X^{j} \epsilon^{a},
\\
\delta F_A^{a} & = - f^{a}_{bc} F_A^{b} \epsilon^{c}
- \frac{\partial f^{a}{}_{bc}}{\partial X^j} F_X^{j} A^b \epsilon^{c}.
\end{align}
In general, $F_A^{a}$ is on-shell covariant.

\section{Off-shell covariantization of 4d algebroid 1-form 
gauge theories}
\label{4d1form}
In the previous sections, we discussed the structure of the Hamiltonian and canonical transformations for general $n$. 
To make the discussion concrete, 
we take a field theory for the specific case $n=3$, i.e. ${\cal M}_{3} = T^{*}[3]E[1]$.
In this case, $\Theta^{(2)}$ can be included in the Hamiltonian function
and we obtain interesting nontrivial examples.

First, we describe the structure of the Hamiltonians based on ${\cal M}_{3}$. 
Local coordinates are $(x^i, q^{a}, \xi_i, p_{a})$ of degree $(0,1,3,2)$, respectively.
Since $\Theta$ is of degree 4, the Hamiltonian function is at most a second order function in $(\xi_i, p_a)$, by degree counting, and can be expanded as
$\Theta = \Theta^{(0)} + \Theta^{(1)} + \Theta^{(2)}$. 
Therefore, the concrete expressions are
\begin{align}
\Theta^{(0)}&=\frac{1}{4!} h_{abcd}(x) q^{a}q^{b}q^{c}q^{d}, 
\label{Thetazero}\\
\Theta^{(1)}&= \frac{1}{2} f^{c}_{ab}(x) q^{a}q^{b}p_{c} 
+ \rho^i{}_a(x)\xi_i q^a,
\label{Thetaone} \\
\Theta^{(2)}&= \frac{1}{2} k^{ab}(x) p_a p_b,
\label{Thetatwo}
\end{align}
with additional structure functions $h_{abcd}(x)$, $f^c{}_{ab}(x)$, 
$\rho^i{}_{a}(x)$ and $k^{ab}(x)$.

From the classical master equation, $\{\Theta,\Theta\}=0$, we 
obtain the following identities,
\begin{eqnarray}
\label{fc1}
&&\rho{}^i{}_{b} k{}^{ba} = 0,\\
\label{fc2}
&&
\rho{}^k{}_{c} \frac{\partial k{}^{ab}}{\partial x^k} 
+ k{}^{da} f{}^b{}_{cd} + k{}^{db} f{}^a{}_{cd} = 0, \\
\label{fc3}
&&
\rho{}^k{}_{b} \frac{\partial \rho{}^i{}_{a}}{\partial x^k} 
- \rho{}^k{}_{a} \frac{\partial \rho{}^i{}_{b}}{\partial x^k} 
+ \rho{}^i{}_{c} f{}^c{}_{ab} = 0,\\
\label{fc4}
&& 
2 \rho{}^k{}_{[d} \frac{\partial f{}^a{}_{bc]}}{\partial x^k} 
+ k{}^{ae} h{}_{bcde}
- 2 f{}^a{}_{e[b} f{}^e{}_{cd]} = 0,\\
\label{fc5}
&& 
2 \rho{}^k{}_{[a} \frac{\partial h{}_{bcde]}}{\partial x^k} 
+ 
\rho{}^f{}_{[ab} h{}_{cde]f} =0,
\end{eqnarray}
which define a Lie 3-algebroid \cite{Ikeda:2010vz}.

Based on the general theory that we explained in the beginning,  
we consider the restriction of the 4-dimensional theory.
The pullback $a^*$ maps the four coordinates to superfields as follows,
\begin{align}
\bcX^{i} &\equiv a^{*}(x^i),\ \bcA^{a} \equiv a^{*}(q^{a}), \\
\bcXi_i & \equiv a^{*}(\xi_i), \ \bcC_a \equiv a^{*}(p_a),
\end{align}
where $(\bcx, \bcA, \bcXi, \bcC)$ are of degree $(0, 1, 3, 2)$.
The super field strengths are given by
\begin{align}
\bcF_X^{i}&= \bcd \bcx^i - \rho^i{}_a(\bcX) \bcA^a,
\label{4superfieldstrength2}
\\
\bcF_A^{a}&= \bcd \bcA^{a} + \frac{1}{2}f^{a}_{bc}(\bcX) \bcA^{b} \bcA^{c}
+ k^{ab}(\bcX) \bcC_b,
\label{4superfieldstrength1}
\\
\bcF^{(C)}_{a}&= \bcd \bcC_{a} + f^{b}_{ac}(\bcX) \bcC_b \bcA^c
- \rho^i{}_a \bcXi_{i}
+ \frac{1}{3!} h_{abcd}(\bcX) \bcA^{b} \bcA^{c} \bcA^{d}, 
\label{4superfieldstrength3}
\\
\bcF^{(\Xi)}_{i}&=\bcd \bcXi_i 
- \frac{1}{2} \frac{\partial f^a{}_{bc}}{\partial \bcX^i}(\bcX)
\bcC_a \bcA^{b} \bcA^{c} 
- \frac{\partial \rho^j{}_{a}}{\partial \bcX^i}(\bcX)
\bcXi_j \bcA^{a}
- \frac{1}{2} \frac{\partial k^{ab}}{\partial \bcX^i}(\bcX)
\bcC_a \bcC_b
\nonumber \\
& + \frac{1}{4!} \frac{\partial h_{abcd}}{\partial \bcX^i}(\bcX) 
\bcA^{a} \bcA^{b} \bcA^{c} \bcA^{d}, 
\label{4superfieldstrength4}
\end{align}
where $\bcF^{(C)}$ and $\bcF^{(\Xi)}$ are the super field strengths of 
$\bcC$ and $\bcXi$, respectively.
When we substitute the component expansions to 
\eqref{4superfieldstrength2}--\eqref{4superfieldstrength4}, then the
corresponding degree $|z|+1$ parts are the field strengths:
\begin{align}
F_X^{i}&= d X^i - \rho^i{}_a(X) A^a,
\label{4fieldstrength1}
\\
F_A^{a}&= d A^{a} + \frac{1}{2}f^{a}_{bc}(X) A^{b} \wedge A^{c}
+ k^{ab}(X) C_b,
\label{4fieldstrength2}
\\
F^{(C)}_{a}&= d C_{a} + f^{b}_{ac}(X) C_b \wedge A^c
- \rho^i{}_a(X) \Xi_{i}
+ \frac{1}{3!} h_{abcd}(X) A^{b} \wedge A^{c} \wedge A^{d}, 
\label{4fieldstrength3}
\\
F^{(\Xi)}_{i}&= d \Xi_i 
- \frac{1}{2} \frac{\partial f^a{}_{bc}}{\partial X^i}(X)
C_a \wedge A^{b} \wedge A^{c} 
- \frac{\partial \rho^j{}_{a}}{\partial X^i}(X)
\Xi_j \wedge A^{a}
- \frac{1}{2} \frac{\partial k^{ab}}{\partial X^i}(X)
C_a \wedge C_b
\nonumber \\ &
+ \frac{1}{4!} \frac{\partial h_{abcd}}{\partial X^i}(X) 
A^{a} \wedge A^{b} \wedge A^{c} \wedge A^{d}.
\label{4fieldstrength4}
\end{align} 
The degree $|z|$ parts of the component expansions of the super field strengths yield the
gauge transformations,
\begin{align}
\delta X^i &= - \rho^i{}_a(X) \epsilon^{a},
\\
\delta A^{a}&= d \epsilon^{a} + f^{a}_{bc}(X) A^{b} \epsilon^{c}
+ k^{ab}(X) \epsilon^{\prime}_b,
\\
\delta C_{a}&= d \epsilon^{\prime}_{a} 
+ f^{b}_{ac}(X) (\epsilon^{\prime}_{b} \wedge A^{c} + C_{b} \wedge \epsilon^{c})
- \rho^i{}_a(X) \mu^{\prime}_i
+ \frac{1}{2} h_{abcd}(X) A^{b} \wedge A^{c} \epsilon^{d}, 
\\
\delta \Xi_{i}&= d \mu^{\prime}_{i} 
- \frac{1}{2} \frac{\partial f^a{}_{bc}}{\partial X^i}(X)
(\epsilon^{\prime}_a \wedge A^{b} \wedge A^{c} + 2 C_a \wedge A^{b} \epsilon^{c})
- \frac{\partial \rho^j{}_{a}}{\partial X^i}(X) 
(\mu^{\prime}_j \wedge A^{a} + \Xi_j \epsilon^{a})
\nonumber \\ &
- \frac{\partial k^{ab}}{\partial X^i}(X) C_a \wedge \epsilon^{\prime}_b
+ \frac{1}{3!} \frac{\partial h_{abcd}}{\partial X^i}(X) 
A^{a} \wedge A^{b} \wedge A^{c} \epsilon^{d}. 
\end{align}
The gauge transformations of the field strengths are
\begin{align}
\delta F_X^{i} &= \frac{\partial \rho^{i}{}_a}{\partial X^j} F_X^{j} \epsilon^{a},
\label{gaugetransformationofFx} \\
\delta F_A^{a} & = - f^{a}_{bc} F_A^{b} \epsilon^{c}
- \frac{\partial k^{ab}}{\partial X^j} F_X^{j} \wedge \epsilon^{\prime}_b
- \frac{\partial f^{a}{}_{bc}}{\partial X^j} F_X^{j} \wedge A^b \epsilon^{c}.
\label{gaugetransformationofFA}
\end{align}
One recognizes from \eqref{gaugetransformationofFA}, that $F_A^a$ does not transform off-shell 
covariantly unless $k^{ab}(X)$ and $f^{a}{}_{bc}(X)$ are constants.

We seek nontrivial deformations of gauge transformations and 
field strengths,
that lead to off-shell covariant gauge structures. 
This is done 
by adding terms to the field strengths and gauge transformations using the fundamental fields 
and lower form field strengths.
Before introducing deformation terms, 
the auxiliary gauge fields are projected out by imposing 
$\Xi_i = C_a =0$.

By form degree counting,
we assume the following structure of deformations of the field strengths
in terms of $X^i$ and $A^a$,
\begin{align}
\hat{F}_X^{i} &= F_X^{i} = d X^i - \rho^i{}_a(X) A^a,
\label{newFx}
\\
\hat{F}_A^{a}&= 
F_A^{a}|_{C_a=0} 
+ K^a_{cj}(X) F_X^{j} \wedge A^{c} + L^a_{ij}(X) F_X^{i} \wedge F_X^{j}
\nonumber \\
& = d A^{a} + \frac{1}{2}f^{a}_{bc}(X) A^{b} \wedge A^{c}
+ K^a_{cj}(X) F_X^{j} \wedge A^{c} + L^a_{ij}(X) F_X^{i} \wedge F_X^{j},
\label{newFA}
\end{align}
where $K^a_{ci}(X)$ and $L^a_{ij}(X)$ are functions.
%
The gauge transformations of $(X^i, A^a)$ should be of the following form,
\begin{align}
\hat{\delta} X^i &= \delta X^i = - \rho^i{}_a(X) \he^{a},
\label{newgaugetransfx}
\\
\hat{\delta} A^{a}&= 
\delta A^{a} + N^a_{ci}(X) F_X^i \he^c
\nonumber \\ 
&=
d \he^{a} + f^{a}_{bc}(X) A^{b} \he^{c}
+ N^a_{ci}(X) F_X^i \he^c,
\label{newgaugetransfA}
\end{align}
where $N^a_{ci}(X)$ is a function.

Let us compute the gauge transformations of \eqref{newFx}
and \eqref{newFA} using  
\eqref{newgaugetransfx} and \eqref{newgaugetransfA}.
Employing the Bianchi identities derived from \eqref{Bianchi},
\begin{align}
d F_X^{i}&= \frac{\partial \rho^i{}_a}{\partial X^j} F_X^j A^a
+ \rho^i{}_a F_A^a,
\end{align}
we can compute $\hat{\delta} F_A^{a}$.
The requirement that the coefficients of $F_X^i d \he^b$, 
$F_X^i \wedge A^a \he^b$, $F_X^i \wedge A^a \he^b$ 
and $F_X^i \wedge F_X^j \he^b$ 
in $\hat{\delta} F_A^a$ vanish gives relations
among $K, L$ and $N$,
\begin{align}
& N^a_{bi} = K^a_{bi},
\label{KLNcondition1}
\\
& \frac{\partial f^{a}{}_{bc}}{\partial x^i} 
+ f^a{}_{db} K^d_{ci} 
+ K^a_{di} f^d{}_{bc} 
- \frac{\partial N^{a}_{ci}}{\partial x^j} \rho^j{}_b
+ \frac{\partial K^{a}_{bi}}{\partial x^j} \rho^j{}_c
- N^a_{cj} \frac{\partial \rho^{j}{}_b}{\partial x^i}
+ K^a_{bj} \frac{\partial \rho^{j}{}_c}{\partial x^i}
= f^a{}_{dc} K^d_{bi},
\label{KLNcondition2}
\\
&  
\frac{1}{2} \frac{\partial N^a_{cj}}{\partial x^i}
+ \frac{1}{2} K^a_{bi} N^b_{cj} 
+ L^a_{ki} \frac{\partial \rho^{k}{}_c}{\partial x^j}
- (i \leftrightarrow j) = f^a{}_{bc} L^b_{ij}
- \frac{\partial L^{a}_{ij}}{\partial x^k} \rho^k{}_c.
\label{KLNcondition3}
\end{align}
Under these conditions the field strength is off-shell covariant,
\begin{align}
\hat{\delta}\hat{F}_A^{a}&= - (f^a{}_{bc} + N^a_{ci} \rho^i{}_b) \hat{F}_A^{b} \he^c.
\end{align}

\subsection{Examples}
\subsubsection{St\"uckelberg formalism}
First, we consider a trivial example to show that this formalism is 
a generalization of a known formalism. The starting point is the QP-manifold 
$\mathcal{M}_3=T^*[3]E[1]$, where  $E=TM$ is a tangent bundle. 
We take
\begin{align}
f^a{}_{bc} &= k^{ab}(x) = h_{abcd} = 0, \notag
\\
\rho^i{}_a &= m \delta^i{}_a = \mbox{constant}, \notag
\end{align}
where $i$ and $a$ run over the same index range. 
Then, the Hamiltonian function is
\begin{align}
\Theta &= m \xi_a q^a,
\end{align}
which trivially satisfies the classical master equation, $\{\Theta, \Theta\}=0$.
The resulting field strengths are
\begin{align}
F_X^{a}&= d X^a - m A^a,
\label{sfieldstrength1}
\\
F_A^{a}&= d A^{a},
\label{sfieldstrength2}
\\
F^{(C)}_{a}&= d C_{a} + m \Xi_a,
\label{sfieldstrength3}
\\
F^{(\Xi)}_{a}&= d \Xi_a.
\label{sfieldstrength4}
\end{align} 
The gauge transformations of the gauge fields are
\begin{align}
\delta X^a &= - m \epsilon^a,
\\
\delta A^{a}&= d \epsilon^{a},
\\
\delta C_{a}&= d \epsilon^{\prime}_{a} + m \mu^{\prime}_a,
\\
\delta \Xi_{a}&= d \mu^{\prime}_{a}.
\end{align}
From these equations, the gauge transformations of the field strengths 
are trivially covariant,
\begin{align}
\delta F_X^{a} &= 0,
\label{sgaugetransformationofFx} \\
\delta F_A^{a} & = 0.
\label{sgaugetransformationofFA}
\end{align}
The gauge invariant action,
\begin{align}
S &= \int {\mathrm tr} (F_A \wedge * F_A)
+ {\mathrm tr} (F_X \wedge * F_X)
\nonumber \\
&= \int F_{A\mu\nu} F_A^{\mu\nu}
+ (\partial_{\mu} X^a - m A_{\mu}^a)(\partial^{\mu} X^a - m A^{\mu a}).
\end{align}
is the so-called St\"uckelberg formalism of the massive vector field $A_{\mu}^a$.
We conclude that our formalism provides a nonlinear generalization of 
the St\"uckelberg formalism.

\subsubsection{Nonabelian gauged nonlinear sigma models}
We list a simple but nontrivial example, taking again $\mathcal{M}_3$ as a starting point.
Let the structure constants be
\begin{align}
f^a{}_{bc}  = \mbox{constant},
\quad 
\rho^i{}_a = h_{abcd} = 0, 
\quad
k^{ab}(x) = \mbox{arbitrary}.\notag
\end{align}
Then, the Hamiltonian function is 
\begin{align}
\Theta &= \frac{1}{2} f^a{}_{bc} q^b q^c p_a
+ \frac{1}{2} k^{ab}(x) p_a p_b.
\end{align}
The resulting field strengths are
\begin{align}
F_X^{i}&= d X^i,
\label{efieldstrength1}
\\
F_A^{a}&= d A^{a} 
+ \frac{1}{2} f^a{}_{bc} A^b A^c
+ k^{ab} C_{b},
\label{efieldstrength2}
\\
F^{(C)}_{a}&= d C_{a}
+ f^b{}_{ac} A^c \wedge C_b,
\label{efieldstrength3}
\\
F^{(\Xi)}_{i}&= d \Xi_i 
- \frac{1}{2} \frac{\partial k^{ab}}{\partial X^i}(X)
C_a \wedge C_b.
\label{efieldstrength4}
\end{align} 
The gauge transformations of the gauge fields are
\begin{align}
\delta X^i &= 0,
\\
\delta A^{a}&= d \epsilon^{a} 
+ f^a{}_{bc} A^b \epsilon^c
+ k^{ab}(X) \epsilon^{\prime}_b,
\\
\delta C_{a}&= d \epsilon^{\prime}_{a}
+ f^b{}_{ac} (A^c \wedge \epsilon^{\prime}_b + \epsilon^c C_b),
\\
\delta \Xi_{i}&= d \mu^{\prime}_{i} 
- \frac{\partial k^{ab}}{\partial X^i}(X) C_a \wedge \epsilon^{\prime}_b.
\end{align}
Using these equations, we compute the gauge transformations of the field strengths 
as
\begin{align}
\delta F_X^{i} &= 0,
\label{egaugetransformationofFx} \\
\delta F_A^{a} & = 
- f^a{}_{bc} F_A^b \epsilon^c
- \frac{\partial k^{ab}}{\partial X^j} F_X^{j} \wedge \epsilon^{\prime}_b.
\label{egaugetransformationofFA}
\end{align}
The gauge transformation of $F_A^{a}$ is not off-shell covariant.

Let us apply our formalism to this system.
A solution of \eqref{KLNcondition1}--\eqref{KLNcondition3} in 
this example is
\begin{align}
K^a_{bi} &= N^a_{bi} = \delta^a{}_b\frac{\partial w}{\partial x^i}(x),
\quad 
L^a_{ij}=0,
\end{align}
where $w(x)$ is an arbitrary function.
The covariantized field strengths and gauge transformations are 
computed as 
\begin{align}
\hat{F}_X^{i}&= F_X^i = d X^i,
\\
\hat{F}_A^{a}&= d A^{a}
+ \frac{1}{2} f^a{}_{bc} A^b A^c
+ \frac{\partial w}{\partial X^i} F_X^{i} \wedge A^{a},
\\
\hat{\delta} A^{a}&= d \he^{a} 
+ f^{a}_{bc} A^{b} \he^{c}
+ \frac{\partial w}{\partial X^i} F_X^{i} \he^a,
\\
\hat{\delta} X^i &= 0.
\end{align}
Finally, we obtain
\begin{align}
\hat{\delta} F_X^{i}&= 0,
\\
\hat{\delta} \hat{F}_A^{a}&= - f^a{}_{bc} \hat{F}_A^b \he^c.
\end{align}

Assume that $M$ is 1-dimensional. Then, we drop the index $i$ and take
\begin{align}
K^a{}_{b} &= N^a{}_{b} = \frac{\delta^a{}_b}{x},
\end{align}
which yields
\begin{align}
F_X&= d X,
\\
\hat{F}_A^{a}&= d A^{a}
+ \frac{1}{2} f^a{}_{bc} A^b A^c + \frac{1}{X} F_X \wedge A^{a},
\\
\hat{\delta} A^{a}&= d \he^{a} 
+ f^{a}_{bc} A^{b} \he^{c}
+ \frac{1}{X} F_X \he^a,
\\
\hat{\delta} X &= 0.
\end{align}
By the redefinition of $x$ via 
\begin{align}
\varphi &= \log |X|,
\end{align}
the equations can be rewritten in a nonsingular form,
\begin{align}
F_x&= e^{\varphi} d \varphi,
\\
\hat{F}_A^{a}&= d A^{a}
+ \frac{1}{2} f^a{}_{bc} A^b A^c + d \varphi \wedge A^{a},
\\
\hat{\delta} A^{a}&= d \he^{a} 
+ f^{a}_{bc} A^{b} \he^{c}
+ d \varphi \he^a,
\\
\hat{\delta} \varphi &= 0.
\end{align}

\subsubsection{Kotov-Strobl model}
As third example we formulate the model proposed in \cite{Kotov:2015nuz}.

Here, we consider a QP-manifold of degree two, ${\cal M}_2 = T^*[2]E[1]$,
in order to demonstrate the covariantization procedure for the Kotov-Strobl model.
Note that the resulting gauge theory is not restricted to any dimension.
The local coordinates of ${\cal M}_2$
are denoted by $(x^i, \xi_i, q^a)$ of degree $(0,2,1)$.
The fiber coordinates of $E[1]$ and $E^*[1]$ are identified by introducing a fiber metric 
$\lambda_{ab}$.\footnote{We can use the fiber coordinates $(q^a, p_a)$ of $E[1]$ and $E^*[1]$.}
The graded symplectic form is defined by
\begin{align}
\omega = \delta x^i \wedge \delta \xi_i + \frac{1}{2} \lambda_{ab}(x) \delta q^a \wedge \delta q^b.
\end{align}
The most general form of the Hamiltonian is given by 
\begin{align}
\Theta = \rho^i{}_a(x) \xi_i q^a + \frac{1}{3!} h_{abc}(x) q^a q^b q^c.
\end{align}

In order to construct the Kotov-Strobl model,
we take $M$ be a 2-dimensional manifold and $E$ a vector bundle over $M$
with 1-dimensional fiber.
Let us denote the local coordinates of ${\cal M}_2$ by
$(x, y):=(x^1, x^2)$, 
$(\xi, \eta):=(\xi_1, \xi_2)$ and $q:=q^1$ and
take the following Hamiltonian function,
\begin{align}
\Theta 
& = - e^{- \frac{\lambda}{2} xy} \eta q,
\label{ThetaKotovStrobl}
\end{align}
where $\lambda$ is a constant. That corresponds to choosing
\begin{align}
h{}_{abc} & 
= 0, 
\ 
\lambda_{11} = 1,
\notag
\\
\rho^1 &= 0, \ 
\rho^2 = e^{- \frac{\lambda}{2} xy}. \notag
\end{align}
The associated superfields are defined as
\begin{align}
\bcx &\equiv a^{*}(x),\  \bcy \equiv a^{*}(y), \notag \\ 
\bcA &\equiv a^{*}(q),
\notag \\
\bcXi & \equiv a^{*}(\xi), \ \bcH \equiv a^{*}(\eta). \notag 
\end{align}
Using the formulas \eqref{generalfieldstrength} and \eqref{superfeildstrength}
, we obtain the following field strengths,
\begin{align}
F_X&= d X,
\label{ksfieldstrength1}
\\
F_Y&= d Y - e^{- \frac{\lambda}{2} XY} A,
\label{ksfieldstrength2}
\\
F_A&= d A + e^{- \frac{\lambda}{2} XY} H,
\label{ksfieldstrength3}
\\
F_{\Xi}&= d \Xi - \frac{\lambda}{2} Y e^{- \frac{\lambda}{2} XY} H A,
\label{ksfieldstrength7}
\\
F_{H} &= d H - \frac{\lambda}{2} X e^{- \frac{\lambda}{2} XY} H A,
\label{ksfieldstrength8}
\end{align} 
and gauge transformations of the gauge fields,
\begin{align}
\delta X &= 0,
\\
\delta Y &= - e^{- \frac{\lambda}{2} XY} \epsilon,
\\
\delta A &= d \epsilon + e^{- \frac{\lambda}{2} XY} \mu^{\prime}_2,
\\
\delta \Xi &= d \mu_1^{\prime} - \frac{\lambda}{2} Y e^{- \frac{\lambda}{2} XY} (\mu^{\prime}_2 A + H \epsilon),
\\
\delta H &= d \mu_2^{\prime} - \frac{\lambda}{2} X e^{- \frac{\lambda}{2} XY} (\mu^{\prime}_2 A + H \epsilon).
\end{align}
Here, $\epsilon$ is the 0-form gauge parameter corresponding to $A$,
and $\mu^{\prime}_1$ and $\mu^{\prime}_2$ are the 1-form gauge parameters corresponding to $\Xi$ and $H$, respectively.
We are only interested in the gauge transformations and field strengths 
of the fields $(X, Y, A)$.
The gauge transformations of the field strengths are computed as
\begin{align}
\delta F_X &= 0,
\label{ksgaugetransformationofFx} \\
\delta F_Y &= 
- \frac{\lambda}{2} Y e^{- \frac{\lambda}{2} XY} F_X \epsilon
- \frac{\lambda}{2} X e^{- \frac{\lambda}{2} XY} F_Y \epsilon
- e^{- \lambda XY} \mu^{\prime}_2,
\label{ksgaugetransformationofFy} \\
\delta F_A^{a} & = 
\frac{\lambda}{2} Y e^{- \frac{\lambda}{2} XY} F_X \mu^{\prime}_2
+ \frac{\lambda}{2} X e^{- \frac{\lambda}{2} XY} F_Y \mu^{\prime}_2.
\label{ksgaugetransformationofFA}
\end{align}
The gauge transformations of $F_Y$ and $F_A$ are not off-shell covariant.

We apply the off-shell covariantization procedure to this theory.
The possible deformations of the field strengths and gauge transformations are
\begin{align}
\hat{F}_A&= F_A + J(X, Y) F_X \wedge A + K(X, Y) F_Y \wedge A 
+ L(X, Y) F_X \wedge F_Y, 
\\
\hat{\delta} A &= d \he + M(X, Y) F_X \wedge \he + N(X, Y) F_Y \wedge \he,
\end{align}
where we determine the functions  $J, K, L, M, N$ of the scalar fields $X, Y$.
Deformations of the other field strengths and gauge transformations need not to be considered.

One solution is $M = - \frac{\lambda}{2} Y$ and $N=0$.
In this case, 
$\hat{\delta} F_Y$ is covariantized as
\begin{align}
\hat{\delta} F_Y &=  
- \frac{\lambda}{2} e^{- \frac{\lambda}{2} XY} X F_Y \he.
\end{align}
In the next step, we require off-shell covariance of 
$\hat{\delta} \hat{F}_A$.
This determines
$J = - \frac{\lambda}{2} Y$, $K=0$ and 
$L = - \frac{\lambda}{2} Y e^{\frac{\lambda}{2} XY}$.
The resulting field strengths and gauge transformations are
\begin{align}
\hat{F}_A &= d A
- \frac{\lambda}{2} Y F_X \wedge A 
- \frac{\lambda}{2} Y e^{\frac{\lambda}{2} XY} F_X \wedge F_Y
\nonumber \\
&= d A - \frac{\lambda}{2} Y e^{\frac{\lambda}{2} XY} dX \wedge dY,
\\
\hat{\delta} A &= d \he
- \frac{\lambda}{2} Y dX \he.
\end{align}
The gauge transformation of $\hat{F}_A$ is computed as
\begin{align}
\hat{\delta} \hat{F}_A &= 0,
\end{align}
which is off-shell covariant.

\paragraph{Invariant Action}
Since the scalar field strength $F_X^i = dX^i - \rho^i{}_a(X) A^a$
transforms off-shell covariantly, 
\begin{align}
\hdel F_X^i 
= \frac{\partial \rho^{i}{}_a}{\partial X^j} F_X^{j} \he^{a},
\end{align}
the action 
\begin{align}
\int g_{ij}(X) F_X^i \wedge * F_X^j
\end{align}
is invariant if $g_{ij}(X)$ satisfies
\begin{align}
\hdel g_{ij}(X) &= - \left(g_{kj} \frac{\partial \rho^k{}_a}{\partial X^i}
+ g_{ik} \frac{\partial \rho^k{}_a}{\partial X^j} \right) \he^a.
\end{align}
In this example, the action is given by
\begin{align}
S &= \int 
F_X \wedge * F_X + V(X) 
+ e^{\lambda XY} F_Y \wedge * F_Y + \hFA \wedge * \hFA
\end{align}
and is invariant under gauge transformations.
The gauge transformation of the third term is given by
\begin{align}
\hdel(e^{\lambda XY} F_Y \wedge * F_Y )
&= 2 e^{\lambda XY} (\hdel F_Y) \wedge * F_Y 
- (\hdel e^{\lambda XY}) F_Y \wedge * F_Y 
\nonumber \\
&= e^{\lambda XY} \lambda X \epsilon F_Y \wedge * F_Y 
- e^{\lambda XY} \lambda X \epsilon F_Y \wedge * F_Y =0.
\end{align}

\paragraph{Correspondence to the Kotov-Strobl model}
By the off-shell covariantization procedure, we obtain the field strengths
\begin{align}
F_X&= d X,
\label{ksfieldstrength1}
\\
F_Y&= d Y - e^{- \frac{\lambda}{2} XY} A,
\label{ksfieldstrength2}
\\
\hFA&= d A + \frac{\lambda}{2} e^{\frac{\lambda}{2} XY} Y d Y \wedge dX,
\label{ksfieldstrength3}
\end{align} 
and the gauge transformations of the gauge fields
\begin{align}
\hdel X &= 0,
\\
\hdel Y &= - e^{- \frac{\lambda}{2} XY} \he,
\\
\hdel A &= d \he - \frac{\lambda}{2} d X Y \he.
\end{align}
From these equations, the gauge transformations of the field strengths 
are computed as
\begin{align}
\hdel F_X &= 0,
\label{ksgaugetransformationofFx} \\
\hdel F_Y &= 
- \frac{\lambda}{2} e^{- \frac{\lambda}{2} XY} X F_Y \he,
\label{ksgaugetransformationofFy} \\
\hdel F_A & = 0.
\label{ksgaugetransformationofFA}
\end{align}
Redefining the gauge field $A$, the gauge parameter $\epsilon$ and 
the field strength $F_A$ as
\begin{align}
\tilde{\epsilon} &\equiv e^{- \frac{\lambda}{2} XY} \he,
\\
\tilde{A} &\equiv e^{- \frac{\lambda}{2} XY} A,
\\
G_A &\equiv e^{- \frac{\lambda}{2} XY} \hFA,
\end{align}
we obtain the following field strengths from eqs.
\eqref{ksfieldstrength1}--\eqref{ksfieldstrength3},
\begin{align}
F_X&= d X,
\label{ksfieldstrength11}
\\
F_Y&= d Y - \tilde{A},
\label{ksfieldstrength12}
\\
G_A &= d \tilde{A} + \frac{\lambda}{2} (X dY \wedge \tilde{A} 
+ Y F_Y \wedge dX).
\label{ksfieldstrength13}
\end{align} 
These are the field strengths discussed in \cite{Kotov:2015nuz}.
We can rewrite the gauge transformations of the gauge fields using $\tilde{\epsilon}$
by
\begin{align}
\hdel X &= 0,
\\
\hdel Y &= - \tilde{\epsilon},
\\
\hdel \tilde{A} &= d \tilde{\epsilon} 
- \frac{\lambda}{2} X F_Y \tilde{\epsilon}.
\end{align}
Then, the gauge transformations of the field strengths are given by
\begin{align}
\hdel F_X &= 0,
\label{ksgaugetransformationofFx} \\
\hdel F_Y &= 
 \frac{\lambda}{2} X F_Y \tilde{\epsilon} ,
\label{ksgaugetransformationofFy} \\
\hdel G_A & = \frac{\lambda}{2} X G_A \tilde{\epsilon}
- \frac{\lambda}{2} X F_Y \wedge F_X \tilde{\epsilon} 
+ \left(\frac{\lambda}{2}\right)^2 X (1-Y) F_Y \wedge F_X \tilde{\epsilon},
\label{ksgaugetransformationofFA}
\end{align}
which are the same expressions as in \cite{Kotov:2015nuz}.

\section{Gauge algebras}
\noindent
Finally, we discuss the closure of the gauge symmetry algebra. For this, we write the gauge transformations as 
\begin{align}
\tilde{\delta} X^i &= - \rho^i{}_a(X) \tilde{\epsilon}^{a},
\\
\tilde{\delta} A^{a}&= d \tilde{\epsilon}^{a} 
+ f^{a}_{bc} A^{b} \tilde{\epsilon}^{c} + N^a_{ci} F_X^i \tilde{\epsilon}^c,
\end{align}
where the gauge parameter $\tilde{\epsilon}^a$ is an ordinary function.
We find, that two gauge transformations $\tilde{\delta}_1$ and $\tilde{\delta}_2$ close to  $\tilde{\delta}_3$ by $[\tilde{\delta}_1, \tilde{\delta}_2] = \tilde{\delta}_3$ with $\tilde{\epsilon}_3^a = f^a_{bc} \tilde{\epsilon}_1^b \tilde{\epsilon}_2^c$, where $\tilde{\delta}_i$ denotes the gauge transformation with respective gauge parameters 
$\tilde{\epsilon}_i$, 
\begin{align}
	[\tilde{\delta}_1, \tilde{\delta}_2] X^i & = 
\tilde{\delta}_3 X^i,
\\
	[\tilde{\delta}_1, \tilde{\delta}_2] A^a & = 
\tilde{\delta}_3 A^a + \Lambda^{a}_{ibc} F_x^i 
\tilde{\epsilon}_1^b \tilde{\epsilon}_2^c,
\end{align}
where
\begin{equation}
\Lambda^a_{ibc}
= - \frac{1}{2} N^a_{di} f^d{}_{bc}
+ f^a{}_{dc} N^d_{bi} 
+ \frac{\partial N^a_{bi}}{\partial x^j} \rho^j{}_c
+ N^a_{cj} \frac{\partial \rho^j{}_b}{\partial x^i}
- (b \leftrightarrow c).
\end{equation}
The gauge transformation of $x^i$ is off-shell closed.
Off-shell closure of the gauge transformation of $A^a$ requires
\begin{equation}
\Lambda^a_{ibc} =0,
\label{offshellclosednesscondtion3}
\end{equation}
which is satisfied in our examples.  

\section{Discussion}
In this paper, we generalized the method to obtain 
off-shell covariant gauge transformations and field strengths of higher gauge theories in \cite{Carow-Watamura:2016nwj} and applied it to a system of algebroid gauge theory with 1-form gauge field and scalars.
We demonstrated off-shell covariantization of a 
gauge theory based on a Lie 2-algebroid and a Lie 3-algebroid.
Recall that the resulting gauge theory is not restricted to any dimension.
For covariantization, we deform field strengths and gauge transformations.
The starting point of this procedure is an on-shell (i.e. ${\bf F}_Z =0$) covariant theory.
Since the gauge transformations and field strengths are deformed proportional to the lower field strengths, they are consistent if the theory is kept on-shell.

There are several directions to develop the approach presented in this paper.
The extension of the method to gauge theories 
with Lie n-algebroid gauge symmetry induced from 
a QP-manifold of degree $n$ is straightforward.
Similar conditions corresponding to \eqref{KLNcondition1}--\eqref{KLNcondition3} can be computed for arbitrary $n$.

Here, we have formulated the Kotov-Strobl model using a QP-manifold of degree two.
However, we can also construct the Kotov-Strobl model from a QP-manifold of degree 
three. For this, a further generalization of the procedure
is necessary.
Another possible application of our method is to investigate multiple M5-brane systems \cite{Witten:1995zh, Witten:2009at}. We can add scalar fields to the analysis conducted
in \cite{Carow-Watamura:2016nwj}.
The procedure in this paper can also be applied to investigate the properties of supergravity in connection with tensor hierarchy.
Furthermore, gauge theoretical formulations of gravity
such as the vielbein formalism or the gauge theory of the Poincar\'e group
can also be treated in this formalism.
It would also be interesting to compare the present formalism with the approach
taken in \cite{Batalin2015r}.
We expect that our approach will shed new light on 
the analysis of such systems.

\section*{Acknowledgments}
The authors would like to thank 
Y. Kaneko
for stimulating discussions and valuable comments.
 M.A.H. is supported by Japanese Government (MONBUKAGAKUSHO) Scholarship.
N.I. and S.W. are supported by the Japan-Belgium Bilateral Joint Research Project of JSPS.




\end{document}